\documentclass[preprint,aps,floats,floatfix,amsmath,nofootinbib,superscriptaddress]{revtex4-1}
\usepackage{graphicx,array,dcolumn}
\usepackage{calc,tabularx, epsfig}
\usepackage{hyperref}

\usepackage{amsmath}
\usepackage{amssymb}
\usepackage{multirow}
\usepackage{xspace}
\usepackage{slashed}
\allowdisplaybreaks[1]
\newlength{\figurewidth}
\newcommand{\beq}{\begin{equation}}
\newcommand{\eeq}{\end{equation}}
\newcommand{\bea}{\begin{eqnarray}}
\newcommand{\eea}{\end{eqnarray}}
\newcommand{\ba}{\begin{array}}
\newcommand{\ea}{\end{array}}
\newcommand{\bg}{\bar{g}}
\newcommand{\mn}{{\mu\nu}}

\newcommand{\pt}{\partial}

\newcommand{\al}{\alpha}
\newcommand{\bt}{\beta}
\newcommand{\g}{\gamma}

\newcommand{\lam}{\lambda}
\newcommand{\Lam}{\Lambda}
\newcommand{\G}{\Gamma}

\newcommand{\OM}{\Omega}

\newcommand{\sg}{\sigma}

\makeatother
\begin{document}
%
% define Title, Author, Address, Preprint#
%%%%%%%%%%%%%%%%%%%%%%%%%%%%%%%%%%%%%%%%%%%%%%%
\title{On the Renormalization Group perspective of $\al$-attractors}
\setlength{\figurewidth}{\columnwidth}
%%%%%%%%%%%%%%%%%%%%%%%%%%%%%%%%%%%%%%%%%%%%%%%
%
\author{Gaurav Narain}
\email{gaunarain@itp.ac.cn}
\affiliation{
Kavli Institute for Theoretical Physics China (KITPC), \\
Key Laboratory of Theoretical Physics,
Institute of Theoretical Physics (ITP), 
Chinese Academy of Sciences (CAS), Beijing 100190, P.R. China.}
%
%
%%%%%%%%%%%%%%%%%%%%%%%%%%%%%%%%%%%%%%%%%%%%%%%
\begin{abstract}
In this short paper we outline a recipe for the reconstruction of $F(R)$
gravity starting from single field inflationary potentials in the Einstein frame. 
For simple potentials one can compute the explicit form of 
$F(R)$, whilst for more involved examples one gets a parametric form of $F(R)$. 
The $F(R)$ reconstruction algorithm is used
to study various examples: power-law $\phi^n$, exponential and $\al$-attractors. 
In each case it is seen that for large $R$ (corresponding to large value of inflaton field),
$F(R) \sim R^2$. For the case of $\al$-attractors $F(R) \sim R^2$ for 
all values of inflaton field (for all values of $R$) as $\al\to0$. For generic inflaton potential 
$V(\phi)$, it is seen that if $V^\prime/V \to0$ (for some $\phi$) 
then the corresponding $F(R) \sim R^2$. 
We then study $\al$-attractors in more detail using non-perturbative renormalisation 
group methods to analyse the reconstructed $F(R)$. 
It is seen that $\al\to0$ is an ultraviolet stable 
fixed point of the renormalisation group trajectories. 
\end{abstract}

\maketitle
%
%%%%%%%%%%%%%%%%%%%%%%%%%%%%%%%%%%%%%%%%%%%%%%%
%

%%%%%%%%%%%%%%%%%%%%%%%%%%%%%%%%%%%%%%%%%%%%%%%
\section{Introduction}
\label{intro}
%%%%%%%%%%%%%%%%%%%%%%%%%%%%%%%%%%%%%%%%%%%%%%%

Receding galaxies showed for the first time that the 
universe we live in is expanding. This was one of the 
first few observation testing the Einstein general-relativity 
and giving it a fundamental status. This expansion can be 
beautifully expressed in terms of Freidmann-Lemaitre-Robertson-Walker (FLRW)
metric which follows from the Einstein's equation under the 
assumption of homogeneity and isotropy. The conformally-flat 
metric evolution in time is dictated entirely by the flow of its 
scale-factor. This flow of scale-factor when extended backward 
in time leads to causally disconnected regions of space-time, 
clashing with the homogeneity and isotropic thermal nature observed 
in the CMB, leading to the famous horizon problem. 
The flatness and monopole problem further intensifies 
the severity of the incompleteness of the theory in the
early universe. 

It is seen that these problems gets resolved if an era 
of exponential expansion existed in the early universe 
\cite{Guth1980,Starobinsky1980,Linde1981}. An existence of such an 
exponential era can further explain large-scale structure. 
The only issue is how can such an era may have started 
leading to exponential expansion? 
Investigation of Einstein field equations shows 
that theoretically either gravity needs to be modified 
at high-energy or there exist additional matter whose 
equation of state is such that it leads to exponential expansion \cite{Linde1983}. 
In the former case one needs to modify the gravity action 
by adding an $R^2$-term (or higher-order terms) to Einstein-Hilbert action 
(where $R$ is the Ricci-scalar of the metric) \cite{Starobinsky1982}.
Such higher-derivative terms leads to accelerated expansion. 
In the later case, it is possible to achieve an accelerated phase 
with a slowly rolling scalar field (also known as inflaton) across a nearly flat potential.
 
There are many models which are trying to systematically 
achieve accelerated expansion in the early epoch of the universe
and are compatible with the PLANCK survey data \cite{Martin2013}. 
These include single field model, multi-field models, modified 
gravity, models inspired form strings, etc. However to compares these 
models with data one needs to reduce them to single field 
model where the action distinctively has Einstein-Hilbert 
gravity term and a scalar-field part with kinetic and potential term.
It is realised that to achieve successful inflation with roughly 
$60$ e-folds it is important that the inflationary potential is very flat. 

Particularly interesting are $F(R)$ models of gravity \cite{Sotiriou2008,DeFelice2010}, 
where Starobinsky inflation falls in a subclass \cite{Starobinsky1980,Starobinsky1982}.
These models in Jordan-frame leads to modified equation of 
motion for the evolution of scale-factor of universe. Under 
appropriate choice of parameters it is possible to achieve 
acceptable accelerated expansion. By conformal transformation this 
models can be cast in to Einstein-Hilbert gravity coupled 
minimally with scalar field, where the potential of scalar-field
depends on the form of $F(R)$ \cite{Maeda1987,Barrow1988}. 
This frame is referred to as Einstein-frame. 
The potential in this frame can be compared 
with the PLANCK survey data and appropriately constrained. One can 
reverse this process and ask about the functional form 
of $F(R)$ for the various single field inflationary potential. 
This process is called $F(R)$ reconstruction
\cite{Rinaldi2014,Bamba2014_1, Bamba2014_2,Pizza2014,Broy2014,Myrzakulov2015}.
In this sense the two approaches 
to inflation: modified gravity and new matter are related 
to each other as one can be transformed into another. 

An interesting models of inflation are cosmological attractors where 
the issue with initial conditions gets resolved nicely. A famous 
cosmological attractor is the $\al$-attractors 
\cite{Kallosh2015,Linde2015,Carrasco2015_1,Carrasco2015_2,
Kallosh2016,Linde2016,Pinhero2017}
(another interesting attractor models are in 
scalar-tensor theories of gravity 
\cite{Kallosh2013,Kallosh2013maa,Elizalde2015,Choudhury2017}). 
The model is written with a non-canonical 
kinetic term along with a potential (which can be arbitrary). Under 
field redefinition the kinetic piece acquires canonical form and the 
potential is modified. The model depends on the parameter $\al$ 
(along with set of parameters appearing in potential). The flatness of 
Einstein-frame potential for large field values (for any $\al$) resolve the 
issues with initial conditions. These kind of $\al$-attractor can be naturally 
embedded in the setting of supergravity to get a well defined consistent 
model of inflation following from a fundamental theory. Therefore they 
become favourable models in the context of inflation. Moreover,
the inflationary predictions of these models are robust under 
quantum corrections to the potential \cite{Fumagalli2016}.
Recently it has been suggested that $\al$-attractors 
may describe dark matter and perhaps even dark energy
giving them an even more wider appeal \cite{Mishra2017}.
The interesting thing in these models is the existence of attractor point $\al\to0$.   

The $F(R)$ reconstruction of $\al$-attractors has been recently done 
to see how the corresponding $F(R)$ function behaves for various 
values of $\al$ \cite{Odintsov2016,Bhattacharya2017,Miranda2017}. 
$F(R)$ reconstruction appears as a simple algorithm 
to determine the form of $F(R)$ for the given Einstein frame 
inflationary potential $V(\phi)$. It turns out, as is shown in this paper, 
that only for simple form of potential one can actually work out the explicit form of 
$F(R)$, while in all other case one gets the expression in parametric from. 
In the case of $\al$-attractor one arrives at a parametric form of 
reconstructed $F(R)$, which is used for further study. 

The aim of this paper is to study the quantum field theory of $\al$-attractors 
and compute the renormalisation group flows of the parameters 
present in the theory, to see whether $\al=0$ can appear as a stable 
gaussian fixed point of the RG trajectories. To achieve this we study the 
RG flows of the reconstructed $F(R)$ via non-perturbative flow equation
\cite{Wetterich1992}. An interesting insight is 
gained if one exploits the non-perturbative renormalisation group 
flows of generic $F(R)$ functions \cite{Machado2007,Codello2007,Codello2008,
Ohta20151,Ohta20152,Falls20161,Falls20162}, which have been computed using 
functional renormalisation group equation in the context of 
asymptotic safety scenario \cite{Niedermaier2006,Percacci2007}. 
In this paper we make use of the
already computed RG flow of generic $F(R)$ 
\cite{Ohta20151,Ohta20152,Falls20161,Falls20162} to analyse the 
behaviour of reconstructed $F(R)$ for the case of $\al$-attractor. As the fixed point 
structure and their stability is independent of the field transformation 
therefore such fixed point analysis will be frame independent.  
We study $\al$-attractor in this setting and compute the beta-function
of its parameters from the RG flow of reconstructed $F(R)$. We 
do the fixed point analysis of the flow equation and see that 
the model has a UV attractive fixed point at $\al=0$ 
satisfying the norms of asymptotic safety scenario. 
This adds a welcoming feature to the $\al$-attractor model.

The outline of paper is follows: section \ref{JtoE} describes the 
algorithm for the reconstruction formalism and show how for 
a given potential of scalar field in Einstein-frame gives the 
corresponding $F(R)$. Section \ref{rgflow} discuss about the 
RG flows in general and how qualitative properties (like 
existence of fixed point and eigenvalues) remains unchanged 
under field transformation. Here we then study the RG 
flow of reconstructed $F(R)$ for the case of $\al$-attractor 
using the non-perturbative flow equation and perform the 
fixed point analysis. Finally conclusions are presented in section \ref{conc}.

%%%%%%%%%%%%%%%%%%%%%%%%%%%%%%%%%%%%%%%%%%%%%%%
\section{$F(R)$ reconstruction}
\label{JtoE}
%%%%%%%%%%%%%%%%%%%%%%%%%%%%%%%%%%%%%%%%%%%%%%%

In this section we will outline the procedure for the 
reconstruction of $F(R)$ for the given potential $V(\phi)$ in the 
Einstein frame. We start by considering the transformation of 
Jordan-frame $F(R)$-gravity action to Einstein-frame 
via conformal scaling. The $F(R)$-gravity action in Jordan-frame is given by
\beq
\label{eq:JDact}
S_{J} = \int {\rm d}^dx \sqrt{-g} F(R) \, , 
\eeq
where $g_\mn$ is the metric in Jordan-frame
while $R$ is the corresponding Ricci scalar
\footnote{
Here we use the signature $\{-,+,+,+\}$. Riemann tensor 
is defined as $R_\mn{}^\rho{}_\sg = \pt_\mu \G_\nu{}^\rho{}_\sg
- \cdots$, while $R_\mn = R_{\al\mu}{}^\al{}_\nu$ and 
$R = g^\mn R_\mn$.}. For generality 
the dimension of space-time is kept arbitrary. Following 
the standard procedure \cite{Maeda1987,Barrow1988},
one can write the $F(R)$ action as a scalar-field 
coupled to gravity. The two theories are related 
by Legendre transform,
\beq
\label{eq:legTR}
S = \int {\rm d}^dx \sqrt{-g} \left[
F(\psi) + F^\prime(\psi) (R-\psi) \right] \, ,
\eeq
where $\psi$ is an auxiliary field whose equation of 
motion is $\psi=R$. This when plugged back in eq. (\ref{eq:legTR})
gives $F(R)$ gravity action. At this point one can do 
conformal transformation $g_\mn = \OM^2 \bg_\mn$
to relate the Jordan frame metric $g_\mn$ with the 
Einstein frame metric $\bg_\mn$. On choosing the 
scale factor $\OM(x)$ such that 
$\OM^{d-2} F^\prime(\psi) = (16\pi G)^{-1}$ is satisfied, then 
one gets the dual theory in the Einstein-frame. The action 
of which is given by,
\bea
\label{eq:EFact}
S_E &=& \int {\rm d}^dx \sqrt{-\bg}
\biggl[
\frac{\bar{R}}{16\pi G} - \frac{1}{2} \pt_\mu \phi \pt^\mu \phi 
- V(\phi) \biggr] \, , 
\eea
where $\bar{R}$ is the Ricci scalar of the Einstein 
frame metric $\bg_\mn$
\footnote{
The notation used here is such that the equation of motion 
following from this action is $\bar{R}_\mn - (1/2) \bg_\mn \bar{R} = 8\pi G T_\mn $.
} and
\beq
\label{eq:Fredef}
\frac{\phi}{M^{(d-2)/2}_p} = \frac{1}{a} \ln \frac{F^\prime(\psi)}{M_P^{d-2}}  \, ,
\hspace{3mm}
V(\phi) = \frac{M_P^d(\psi F^\prime - F)}{(F^\prime)^{d/(d-2)}} \, ,
\eeq
where $M^{d-2}_p =(16 \pi G)^{-1}$ is the reduced Planck mass
in the $d$-dimensions and $a=\sqrt{(d-2)/(2(d-1))}$. These equations gives the functional 
form of the potential $V(\phi)$ in the Einstein frame for the given 
$F(R)$ gravity. But it is interesting to ask whether this process 
can be reversed, in the sense that for a given potential $V(\phi)$ 
in the Einstein frame what is the corresponding function
$F(R)$? This reverse process is commonly called
$F(R)$-reconstruction \cite{Rinaldi2014,Bamba2014_1, 
Bamba2014_2,Pizza2014,Broy2014,Myrzakulov2015}. It turns 
out that although this reconstruction recipe sounds like a 
simple algorithm to implement, in practice it is not always 
possible to compute the explicit functional form of $F(R)$. 
In those cases one has to rely on numerical methods or
express it in parametric form where both $F$ and 
$R$ are functions of $\phi$. This complexity arises due to the non-linear 
coupled differential equations involved in the process. Below 
we outline the algorithm for the reconstruction process.

We first express $F^\prime(\psi)$ in terms of 
$\phi$ using the first equation in (\ref{eq:Fredef}). 
This gives
\beq
\label{eq:Fprime}
F^\prime = M_P^{d-2} \exp\left(\frac{a\phi}{M_P^{(d-2)/2}}\right)\, .
\eeq
Taking a derivative of this equation with respect to 
$\psi$ gives,
\beq
\label{eq:Fpp}
F^{\prime\prime} = a M_P^{(d-2)/2} \exp\left(\frac{a\phi}{M_P^{(d-2)/2}} \right)
\frac{{\rm d} \phi}{{\rm d} \psi} \, .
\eeq
Plugging the expression of $F^\prime$ from eq. (\ref{eq:Fprime})
in the equation for $V$ given in eq. (\ref{eq:Fredef}) 
and taking the derivative of this residual 
equation with respect to $\psi$ leads to an expression from which one can 
extract the relation between $\psi$ and $\phi$. For this one will have to
use eq. (\ref{eq:Fpp}). It should be noted that during the Legendre 
transform which is given in eq. (\ref{eq:legTR}), the 
equations of motion for $\psi$ give $\psi=R$
(where $R$ is the Ricci scalar of the metric in Jordan frame). 
This implies that the expression for Jordan frame 
$R$ in terms of $\phi$ is following,
\beq
\label{eq:psi}
aR = M_P^{-(d-2)/2}\exp\left(\frac{2a\phi/(d-2)}{M_P^{(d-2)/2}}\right)
\left[V^\prime(\phi) + \frac{a d V(\phi)}{(d-2)M_P^{(d-2)/2}} \right] \, .
\eeq
One can now make use of the second equation in
(\ref{eq:Fredef}), where one can plug the expression for 
$F^\prime$ and $\psi$ from eq. (\ref{eq:Fprime}) and 
(\ref{eq:psi}) respectively. This will directly give the 
expression for $F$ in terms of $\phi$. 
\beq
\label{eq:Fint}
F(\phi) = \frac{M_P^{(d-2)/2}}{a} \exp\left(\frac{d a \phi/(d-2)}{M_P^{(d-2)/2}}\right)
\left[
V^\prime + \frac{2 a}{(d-2) M_P^{(d-2)/2}} V
\right]  \, .
\eeq
From eq. (\ref{eq:psi} \& \ref{eq:Fint}) it is seen 
that it is a parametric representation of the 
reconstructed $F(R)$. In the case when $V$ is simple 
then it is possible to invert eq. (\ref{eq:psi}) to express 
$\phi$ in terms of $R$, which is then plugged in 
eq. (\ref{eq:Fint}) to get an explicit form for $F(R)$. 
However in most case of interests (for inflationary potentials)
such an inversion is not possible and an 
explicit expression for $F(R)$ cannot be given. In those 
cases one can still write a parametric form and use it 
for further analysis. 

In the following we will work out the functional form 
of $F(R)$ for some simple inflationary potentials in $d=4$.

%%%%%%%%%%%%%%%%%%%%%%%%%%%%%%%%%%%%%%%%%%%%%%%
\subsection{$\phi^n$ potential}
\label{linpot}
%%%%%%%%%%%%%%%%%%%%%%%%%%%%%%%%%%%%%%%%%%%%%%%

In this case $V(\phi) = \lam\phi^n$. This is a very simple form of potential
with just one term. For this potential the eq. (\ref{eq:Fint}) 
and (\ref{eq:psi}) becomes following,
\bea
\label{eq:linpotPAR}
R &=& \frac{\lam \phi^{n-1} \exp(\phi/(\sqrt{3}M_P))}{M_P} 
\left(n\sqrt{3} + \frac{2\phi}{M_P}\right) \, ,
\notag \\
F &=& \lam \phi^{n-1} M_P \exp(2\phi/(\sqrt{3}M_P)) 
\left(n\sqrt{3}+ \frac{\phi}{M_P}\right)  \, .
\eea
It is seen that even for this simple potential it is difficult to 
invert for arbitrary $n$ and express $\phi$ in terms of $R$ 
without involving complicated functions. However the 
parametric form is good and contains same information. 
Specialising for the case of $n=1$ it is realised that 
in this particular case one can actually 
write an explicit form for the function $F(R)$ by rescaling
$\phi \to \bar{\phi} = (\sqrt{3}+2\phi/M_P)/2$
and $R\to \bar{R} = (M_p \sqrt{e} /(2\lam\sqrt{3})) R$. Under 
such a rescaling it is possible to do inversion in the 
first equation in (\ref{eq:linpotPAR}) giving,
\beq
\label{eq:linLam}
\bar{R} = \bar{\phi} e^{\bar{\phi}} \, .
\eeq
This is the equation of the Lambert W-function. 
In terms of Lambert W-function 
one can then express the functional form for $F(R)$ as
\beq
\label{eq:FR_lam}
F(R) = \frac{M_P^3 R^2}{4\lam\sqrt{3}} \left[
\frac{1}{2}\bar{\phi}^{-2} + \bar{\phi}^{-1}
\right] \, ,
\eeq
where $\bar{\phi} = W[(M_P \sqrt{e}/(2\lam\sqrt{3})) R]$. For 
$R \ll (\lam/M_p)$ one can do a series expansion of the $F(R)$ 
function to get,
\beq
\label{eq:FRexp}
F(R) = \frac{M_P}{\sqrt{e}} \biggl[
2 \Lam + M_P R
+ \frac{M_P^2 R^2}{16\Lam} + \cdots
\biggr] \, ,
\eeq
where $\Lam = \lam /4\sqrt{3e}$. 
This has a more familiar appearance where the various terms 
are successive powers of $R$ suppressed by powers of 
$\Lam/M_p$. The large $\phi$ correspond to large $R$-regime. 
It is noticed that for large $\phi$ in $n=1$ case, the exponential 
factor leads over the linear $\phi$ term giving $F(R) \sim R^2$. 

\begin{figure}
\centerline{
\vspace{0pt}
\centering
\includegraphics[width=3.5in,height=4in]{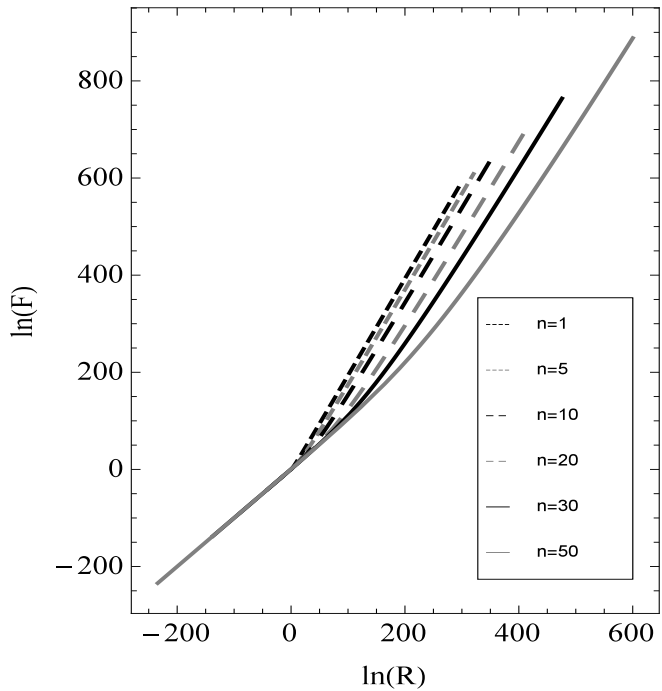}
}
 \caption[]{
$F(R)$ for $\phi^n$. Here on x-axis we plot $\ln(R)$ and 
on y-axis $\ln(F)$. It is seen that for small $R$ the $F(R)\sim R$,
while $F(R) \sim R^2$ for large $R$.  Here we have kept $M_P=1$. 
}
\label{fig:phiN}
\end{figure}

In the more generic case of arbitrary $n$ it is no longer possible to 
invert the first equation of (\ref{eq:linpotPAR}) to express $\phi$ 
in terms of $R$. One is then left with a parametric form of 
$F(R)$. In this case again we note that for small ($\phi \ll M_P$),
we have small $R$. In this regime $F(R) \sim R$. In case of large 
$R$ ($\phi \gg M_P$), one gets $R \sim \lam \phi^n \exp(\phi/(\sqrt{3} M_P))$
and $F \sim \lam \phi^n \exp(2\phi/(\sqrt{3} M_P))$. Here one can see
with some manipulation that $F(R) \sim R^2$ for large $R$. 
Also one can compute the following quantity $d \ln(F)/d \ln(R)$
which gives the rate of variation of $F$ {\it w.r.t.} $R$. In the 
parametric case it is given by,
\beq
\label{eq:parDerphiN}
\frac{{\rm d} \ln(F)}{{\rm d}\ln(R)}
= \frac{{\rm d} \ln(F)}{{\rm d} \phi} \frac{{\rm d} \phi}{{\rm d}\ln(R)}
= \frac{n M_P\sqrt{3} +2 \phi}{n M_P\sqrt{3}  + \phi} \, .
\eeq
From here it is clearly seen that for 
large $\phi$, the {\it r.h.s.} approaches $2$ implying 
$F \sim R^2$. This also signifies the the fact that at high energy (for large $\phi$)
one always approaches a $R^2$ action of gravity. This is like 
an attractor behaviour. These findings can be more clearly seen 
by plotting the parametric reconstructed $F(R)$ as a function 
of $R$ for various $n$. This is shown in figure \ref{fig:phiN}.
%
%%%%%%%%%%%%%%%%%%%%%%%%%%%%%%%%%%%%%%%%%%%%%%%%%%%%%%%
%%%%%%%%%%%%%%%%%%%%%%%%%%%%%%%%%%%%%%%%%%%%%%%%%%%%%%%

%%%%%%%%%%%%%%%%%%%%%%%%%%%%%%%%%%%%%%%%%%%%%%%
\subsection{Exponential potential}
\label{expo}
%%%%%%%%%%%%%%%%%%%%%%%%%%%%%%%%%%%%%%%%%%%%%%%

The other simple scenario is the case when $V(\phi)$ has an 
exponential form. Lets say $V(\phi) = \Lam \exp(b\phi/M_P)$. 
Here $\Lam$ has mass dimension $4$, while $b$ is dimensionless. 
Then using eq. (\ref{eq:psi}) one can express $\phi$ easily in terms of 
$\psi$. This is given by,
\beq
\label{eq:exppot_phi}
\frac{\phi}{M_P} = \frac{\sqrt{3}}{1+b\sqrt{3}} 
\ln \left(\frac{M_P^2 R}{(2+b\sqrt{3})\Lam }\right) \, .
\eeq
Plugging this back in second equation (\ref{eq:Fint}) 
and following the previous foot-steps leads to 
\beq
\label{eq:exppot_fr}
F(R) = (1+b\sqrt{3})\Lam 
\left(\frac{M_P^2 R}{(2+b\sqrt{3})\Lam} \right)^{(2+b\sqrt{3})/(1+b\sqrt{3})}  \, ,
\eeq
In the special case when $b=0$ one gets $R^2$ action. 
If $b=-(n-2)a/(n-1)$, then $F(R) \sim R^n$.  
In the case when $V(\phi) = \Lam (1 -b_1 \exp(b \phi/M_P))^n$, it is 
noticed that 
\bea
\label{eq:stabn1}
R &=& \frac{\Lam}{M_P^2} \exp\left(\frac{\phi}{M_P\sqrt{3}}\right)
(1 -b_1 \exp(b \phi/M_P))^{n-1} 
[ 2 - b_1 (n b\sqrt{3} + 2) \exp(b\phi/M_P)] \, ,
\\
\label{eq:stabn2}
F &=& \Lam \exp\left(\frac{2\phi}{M_P\sqrt{3}}\right)
(1 -b_1 \exp(b \phi/M_P))^{n-1} 
[1 - b_1 (n b\sqrt{3} + 1) \exp(b\phi/M_P)] \, .
\eea
Here it is seen that for arbitrary $n$, $b$ and $b_1$ it is not 
possible to do the inversion and express $\phi$ in terms of $R$.
But for the special case $b = -2/(n\sqrt{3})$ cancellation in the 
expression of $R$ occurs which simplifies the expression of $R$.
This allows one to do the inversion and express $\phi$ in terms of $R$. 
Furthermore if $n=2$ the situation simplifies very much 
and one gets a simple expression for $F(R)$ to be
\beq
\label{eq:staroExp}
F(R) = M_P^2 \left(b_1 R + \frac{M_P^2}{4\Lam} R^2 \right) \, . 
\eeq
This is the usual Starobinsky $R^2$ model which follows 
from the reconstruction process. 

%%%%%%%%%%%%%%%%%%%%%%%%%%%%%%%%%%%%%%%%%%%%%%%
\subsection{$\al$-attractor}
\label{alpha}
%%%%%%%%%%%%%%%%%%%%%%%%%%%%%%%%%%%%%%%%%%%%%%%

\begin{figure}
\centerline{
\vspace{0pt}
\centering
\includegraphics[width=3.5in,height=4in]{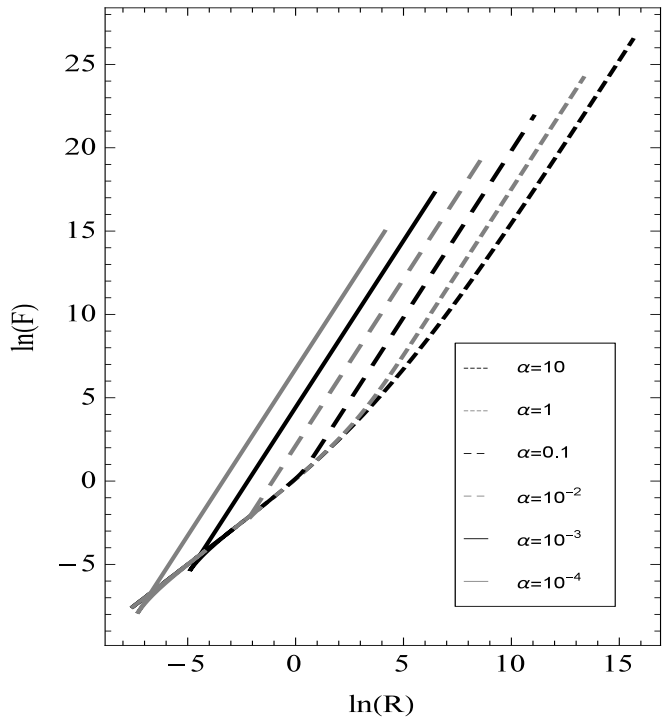}
}
 \caption[]{
Reconstructed $F(R)$ for $\al$-attractor for the $T$-models. 
Here on x-axis we plot $\ln(R)$ and 
on y-axis $\ln(F)$. It is seen that for small $R$ the $F(R)\sim R$,
while for large $R$ $F(R) \sim R^2$. Also as $\al$ approaches 
smaller values, $F \to R^2$. 
}
\label{fig:alphaT}
\end{figure}

An interesting case which is quite famous in context of 
inflation is $\al$-attractors. These were first suggested 
in \cite{Kallosh2015,Linde2015,Carrasco2015_1,Carrasco2015_2} 
which can be easily embedded in supergravity giving it a more 
fundamental status. Here the action has 
a non-canonical kinetic term of the scalar field with a $\phi^2$ type 
potential. The action for $T$-models is given by,
\beq
\label{eq:Actalpha}
S_T = \int {\rm d}^4x \sqrt{-g}
\biggl[
\frac{R}{16\pi G}  - \frac{\pt_\mu \varphi \pt^\mu\varphi}{1 - \varphi^2/6\al}
- \frac{m^2 \varphi^2}{2} 
\biggr] \, ,
\eeq
where $m^2$ and $\al$ are parameters of the theory. 
Under the field transformation $\varphi = \sqrt{6\al} \tanh (\phi/\sqrt{6\al})$
the action acquires a canonical kinetic form where the Einstein 
frame potential is given by,
\beq
\label{eq:VTalpha}
V_T(\phi) = 3 \al m^2 \tanh^2 \left(
\frac{\phi}{\sqrt{6\al}}
\right) \, .
\eeq
Another model of $\al$-attractors is the $E$-models.
Its potential in Einstein frame is,
\beq
\label{eq:VEalpha}
V_E(\phi) = 3 \al m^2 \left(
1 - e^{-\sqrt{2/(3\al)} \phi}
\right)^2 \, .
\eeq
However in the rest of paper we will study only $T$-models 
as the results obtained can be easily extended to $E$-models. 
For $\al$-attractors the potential is complicated function of $\phi$, 
so invertibility during the reconstruction mechanism is not 
possible. However one can still write the parametric form 
of $F(R)$ using the eq. (\ref{eq:psi}) and (\ref{eq:Fint}).
The $\al$-attractor has a special point at $\al=0$. Each 
non-zero value of $\al$ correspond to an inflationary model and acquires a 
point on the $n_s$-$r$ (where $n_s$ is the spectral index
and $r$ is the tensor-scalar ratio). When $\al\to0$ smoothly then 
on the $n_s$-$r$ plane the corresponding point is seen to approach smoothly 
a fixed point, at which $r=0$ and $n_s = 1-2/N$ (where $N$ is the 
number of $e$-folds). 
For small $\al$ the model behaves very much like 
pure $R^2$ inflationary model. In fact if one does the 
reconstruction and plots $F(R)$ for various values of 
$\al$ (see \cite{Odintsov2016,Bhattacharya2017,Miranda2017}
for earlier work on $F(R)$ reconstruction of $\al$-attractor), 
then it is seen that for small $\al$, $F(R)\sim R^2$. 
If we compute the derivative $d \ln(F)/d \ln(R)$ 
then it is given by,
\beq
\label{eq:parDerphiN}
\frac{{\rm d} \ln(F)}{{\rm d}\ln(R)}
= \frac{2\sqrt{2} M_P+2\sqrt{\al} \sinh(\sqrt{2/3\al}\phi)}
{2\sqrt{2} M_P + \sqrt{\al} \sinh(\sqrt{2/3\al}\phi) } \, .
\eeq
Keeping $\al$ fixed it is seen that there are two distinct phases:
small $\phi$ where $F(R) \sim R$ and large $\phi$ where 
$F(R) \sim R^2$. By varying $\al$ it is seen that as $\al\to0$
then this derivative approaches $2$ for all $\phi$, indicating 
that $\al\to0$ is a $R^2$ gravity. These findings can also 
be inferred by plotting the reconstructed $F(R)$ for 
the case of $T$-models (for $E$-models it is similar) 
for various values of $\al$ for a range of values of $\phi$. 
These are presented in figure \ref{fig:alphaT}. 
%
%%%%%%%%%%%%%%%%%%%%%%%%%%%%%%%%%%%%%%%%%%%%%%%%%%%%%%%
%%%%%%%%%%%%%%%%%%%%%%%%%%%%%%%%%%%%%%%%%%%%%%%%%%%%%%%

This is good indication of $R^2$ being an attractor point 
at high energies. This classical analysis motivates 
us to indulge in to investigating whether 
$\al=0$ can be a stable fixed point in the corresponding quantum theory,
where $\al=0$ arises as the ultraviolet stable fixed point 
of the renormalisation group trajectory. If such a thing is possible 
then it gives different perspective to the $\al=0$ point. 
In the next section therefore we 
will study the quantum theory of the $\al$-attractors and compute the 
RG flow of the coupling parameters.

%%%%%%%%%%%%%%%%%%%%%%%%%%%%%%%%%%%%%%%%%%%%%%%
\subsection{Generic $V(\phi)$}
\label{genV}
%%%%%%%%%%%%%%%%%%%%%%%%%%%%%%%%%%%%%%%%%%%%%%%

In the case when the potential is generic $V(\phi)$ then using 
eq. (\ref{eq:psi} \& \ref{eq:Fint}) one can write reconstructed 
$F(R)$. In generic case one can take derivative $d \ln(F)/d \ln(R)$ 
to see the form of the resulting expression. This is given by,
\beq
\label{eq:genVder}
\frac{{\rm d} \ln(F)}{{\rm d}\ln(R)}
= \frac{2\sqrt{3} V(\phi) + 3 M_P V^\prime(\phi)}
{\sqrt{3} V(\phi) +  3 M_P V^\prime(\phi)} \, .
\eeq
This simple expression shows that if $V^\prime/V$
vanishes then one gets the derivative to be $2$, indicating 
$R^2$ gravity. For the forms of $V$ considered above it is 
seen that $V^\prime/V$ vanishes for large $\phi$:
for polynomial forms of potential, $\al$-attractor, 
power-law. In the case of exponential potential 
this ratio vanishes only for special value of $b$ 
(the coefficient appearing the exponential). In case of 
$\al$-attractor this ratio vanishes for all values of $\phi$
as $\al\to0$, giving $R^2$ gravity. 

%%%%%%%%%%%%%%%%%%%%%%%%%%%%%%%%%%%%%%%%%%%%%%%
\section{RG flows}
\label{rgflow}
%%%%%%%%%%%%%%%%%%%%%%%%%%%%%%%%%%%%%%%%%%%%%%%

As the point $\al\to0$ hold a special status in the studies of cosmological attractor, 
one wonders if such a point can have additional interesting qualities. 
In particular one wonders if one can constructs a
quantum theory of such attractor models and study the 
renormalisation group flows of the various parameters.
In this context it is interesting to look for fixed points 
of the RG trajectory, in particular to see whether $\al=0$ 
is a stable fixed point of the RG trajectories. This is the question 
we seek to answer in this short paper {\it i.e.} looking 
at the $\al\to0$ limit from a RG perspective. 

For a generic quantum field theory the action can be written as
\beq
\label{eq:genact}
\G = \sum_i g_i {\cal O}_i \, ,
\eeq 
where $g_i$ are energy dependent couplings while ${\cal O}_i$ 
are the operators. The corresponding beta-functional 
(set of beta-functions of various couplings) is given by 
\beq
\label{eq:betagen}
\pt_t \G = \sum_i \bt_i {\cal O}_i 
= (\pt_t g_i) {\cal O}_i\, ,
\eeq
where $t= \ln (\mu/\mu_0)$ is the RG time and $\mu$ is the 
running energy scale, while $\mu_0$ is the reference point. 
A beta-functional is a vector field on the space of 
couplings. The fixed points on this coupling space correspond 
to points which are solutions of $\bt_i=0$ (for all $i$). The behaviour of flows around the 
fixed points tell us about the stability of fixed points. If the fixed point
is given by $g^*_i$, then near the fixed point the flow will acquire a 
linearised form given by,
\beq
\label{eq:linflow}
\pt_t g_i=  \sum_j M_{ij} (g_j - g^*_j) \, ,
\eeq
where 
\beq
\label{eq:linmat}
M_{ij} = \left. \frac{\pt \bt_i}{\pt g_j} \right|_{g=g*} \, ,
\eeq
is a linearised matrix computed at the fixed point
(also known as stability matrix). The nature of 
eigenvalues dictates the stability of the fixed point. Corresponding to 
each eigenvalue is an eigenvector, which form the 
basis of the eigenspace around the fixed point. If the eigenvalue 
is negative then corresponding eigen-direction is UV attractive (infrared repulsive) 
in nature, while positive eigenvalue correspond to UV repulsive (and IR attractive) 
eigen-direction. For zero eigenvalues one has to do second order analysis 
in order to determine the true nature of the corresponding 
eigenvector. The interesting thing to note here is that 
while beta-functional behaves as a vector-field in the 
coupling space, the quantity $M_{ij}$ behaves like a 
tensor at the fixed point. This becomes more clear when a field transformation is done.
In such a case the new couplings become functions of older ones, 
while the beta-function of new couplings $\bar{g}_i$ 
are related to beta-function of old couplings $g_i$ via 
vector-transformation in coupling space as
\beq
\label{eq:betaTR}
\pt_t \bar{g}_i = \frac{\pt \bar{g}_i}{\pt g_j} \pt_t g_j \, ,
\eeq
where $\pt \bar{g}_i/\pt g_j$ is the Jacobian of transformation. 
This will imply that if a certain point in RG trajectory is a fixed 
point, then the corresponding point in the transformed space 
is also a fixed point. Under such a transformation the stability matrix
$M_{ij}$ transform as,
\beq
\label{eq:MTR}
\bar{M}_{ij} = \frac{\pt g_k}{\pt \bar{g}_j} 
\frac{\pt^2 \bar{g}_i}{\pt g_k \pt g_l} (\pt_t g_i)
+ \frac{\pt g_k}{\pt \bar{g}_j} \frac{\pt \bar{g}_i}{\pt g_l} M_{lk} \, .
\eeq
At the fixed point the first term on the RHS vanishes as the beta-functions
are zero, thereby leaving only a tensor transformation for $M_{ij}$. Such 
tensor transformation leave the eigenvalues of the stability matrix invariant, 
but eigenvector gets rotated. 

In the case of $\al$-attractor the reconstructed $F(R)$ gravity action is the 
theory in Jordan frame and is related to the $\al$-attractor model via 
a non-linear field transformation. It has been still a debatable issue whether the 
quantum theories in the two frames in general are same or more so whether 
one should only construct quantum theories in Jordan frame. 
There has been some examples of scalar-tensor theories of gravity 
where quantum equivalence has been studied 
\cite{Kamenshchik2014,Banerjee2016,Pandey2016},
and it has been seen that the quantum theories in the two frames are same. 
In the case of $F(R)$ theories although such a study is currently lacking, but 
under field transformations while moving from Jordan to Einstein frame, it 
is realised that a non-minimally coupled scalar-tensor theory emerges at an intermediate 
level. This is like the ones considered in \cite{Kamenshchik2014,Banerjee2016,Pandey2016},
whose equivalence to Einstein frame theory has been shown. 
Therefore one can argue with good reasons that 
a quantum equivalence for $F(R)$ theory will also exist. 
In this paper we exploit this knowledge to get a different 
perspective on the $\al$-attractor using the non-perturbative 
RG flows of the reconstructed $F(R)$. 

Non-perturbative RG flows of generic $F(R)$ have been computed in 
the past \cite{Ohta20151,Ohta20152,Falls20161,Falls20162}, here we make 
use of those generic RG equation in our case to extract information 
about the running of parameters of $\al$-attractors. The equation 
was written on the background of maximally symmetric spaces 
for generic gauge-fixing. This generality works in our favour 
and can be specialised in our case. However their equation 
was written for an explicit function of $F(R)$. As we know that 
in most cases of reconstruction it is not possible to 
give an explicit form of $F(R)$ due to difficulty in invertibility, 
but it is possible to have a parametric form of $F(R)$ where 
both $F$ and $R$ are functions of $\phi$ given by 
eq. (\ref{eq:psi}) and (\ref{eq:Fint}) respectively. Then one need to 
adapt the RG equation presented in \cite{Ohta20151,Ohta20152,Falls20161,Falls20162}
for the explicit $F(R)$ to the parametric case.
This adaptation is achieved by making use of chain-rule of derivatives. 

On maximally symmetric spaces the Riemann curvature tensor and Ricci tensor 
satisfy 
\beq
\label{eq:Reim}
R_{\mn\rho\sg} = \frac{R}{d(d-1)} (g_{\mu\rho}g_{\nu\sg}
- g_{\mu\sg}g_{\nu\rho}) \, ,
\hspace{5mm}
R_\mn = \frac{R}{d} g_\mn \, ,
\eeq
where $R$ doesn't depend on space-time. 
In the case of $F(R)$ reconstruction, $R$ is function of $\phi$, 
which on maximally symmetric spaces remains space-time independent. 
The crucial thing to note now is that any derivative of $F(R)$ 
with respect to $R$ translates into a parametric derivative
\beq
\label{eq:frder}
F^\prime(R) = \frac{{\rm d}F}{{\rm d} \phi} \frac{{\rm d} \phi}{{\rm d} R} \, . 
\eeq
The non-perturbative flow equation for $F(R)$ 
that is written in \cite{Falls20161,Falls20162} contains 
$F^\prime$, $F^{\prime\prime}$ and $F^{\prime\prime\prime}$
in the flow equations. These get accordingly translated to parametric case 
via chain rule. The flow equation in the parametric cases therefore remains 
dependent on parameter $\phi$, where the dependence on $R$ 
comes implicitly. This functional equation is written in dimensionless form, 
where the dimensionless variables are obtained using the RG running 
scale $\mu$. In the parametric case we have,
\beq
\label{eq:dimless}
\bar{\phi} = \mu^{-(d-2)/2} \phi \, , 
\hspace{2mm}
r(\bar{\phi}) = \mu^{-2} R(\phi) \, ,
\hspace{2mm}
\bar{F}(\bar{\phi}) = \mu^{-d} F(\phi).
\eeq
The coupling parameters appearing inside the potential $V(\phi)$
should be accordingly translated to dimensionless form. 
In the present case of $\al$-attractor, the parameter $\al$
has mass dimension two, while the parameter
$m$ has mass-dimension one. These parameters have an
RG running which is extracted from the non-perturbative 
flow equation of the reconstructed $F(R)$. The dimensionless 
parameters are defined as, 
\beq
\label{eq:dimlessPa}
\bar{\al}(t) = \mu^{-(d-2)} \al(t) \, ,
\hspace{2mm}
\bar{m}^2(t) = \mu^{-2} m^2(t) \, .
\eeq
The dimensionless functional RG over the background of 
deSitter space is given by \cite{Ohta20151,Ohta20152}
(for RG flow equation on hyperbolic spaces see \cite{Falls20162}),
\beq
\label{eq:frgFR}
\dot{\bar{F}} - 2 r \bar{F}^\prime + 4 \bar{F}
= \frac{c_1(\dot{\bar{F}}^\prime - 2 r \bar{F}^{\prime\prime}) + c_2 \bar{F}^\prime}
{\bar{F}^\prime(6 + (6\g_1+1)r)}
+ \frac{c_3(\dot{\bar{F}}^{\prime\prime} - 2 r \bar{F}^{\prime\prime\prime}) 
+ c_4 \bar{F}^{\prime\prime}}{\bar{F}^\prime+(3 + (3\g_2-1)r)\bar{F}^{\prime\prime}}
- \frac{c_5}{4+(4\g_3 -1)r} \, ,
\eeq
where $(\dot{})$ denotes derivative {\it w.r.t.} $t$, $({}^\prime)$ denotes implicit 
derivative {\it w.r.t.} $r(\bar{\phi})$. Here $\g_1$, $\g_2$ and $\g_3$ 
are endomorphism parameters which contain information of gauge-fixing.
$c_1$, $c_2$, $c_3$, $c_4$ and $c_5$ are functions of 
$r(\bar{\phi})$ which have been given in the appendix (\ref{ci}). 

In the case of an explicit $F(R)$ function one can obtain the RG running of 
various parameters by projecting the flow equation on various operators. 
This is the usual strategy in the case when $F(R)$ has a series expansion 
in powers of $R$. In the case of implicit form (the parametric case), one 
can still employ similar strategy. It is evident that the resulting dimensionless 
flow equation given in eq. (\ref{eq:frgFR}) is a function of the dimensionless 
parameter $\bar{\phi}$. Taking successive derivatives with respect to $r$ 
and putting $r=0$ is equivalent to taking successive derivatives {\it w.r.t.} $\bar{\phi}$
on both sides of flow equation (\ref{eq:frgFR}) and putting 
$\bar{\phi}=0$. This is possible due to chain rule 
${\rm d}/{\rm d} r = ({\rm d} \bar{\phi}/ {\rm d} r) {\rm d}/{\rm d} \bar{\phi}$,
where ${\rm d} r/ {\rm d} \bar{\phi} \neq 0$ at the $\bar{\phi}=0$. 
This allow us to extract the running of parameter of the theory. 
In the case of $\al$-attractors it is seen 
that to extract the running of $\bar{m}^2(t)$ and $\bar{\al}(t)$ one has to 
do series expansion in $\bar{\phi}$ of both sides of flow equation.
Here we did the expansion up to $\bar{\phi}^3$. 
This is sufficient to extract the running of parameters. However 
in the functional RG framework this is a truncation, and the flow equation 
of these parameters will be truncation dependent. For example the 
running of parameters $\bar{m}^2(t)$ and $\bar{\al}(t)$ can also be 
extracted from the coefficient of $\bar{\phi}^3$ and $\bar{\phi}^4$
or $\bar{\phi}^4$ and $\bar{\phi}^5$, etc. These are higher-order truncations. 
In these cases the flow equations
obtained will be more complicated with lengthy expression. However 
certain features of the flow equations will be independent of truncations
for example the existence of non-spurious fixed points. 

The flow equations for the dimensionless 
parameters $\bar{\al}(t)$ and $\bar{m}^2(t)$ have the following form,
\bea
&&
\label{eq:floweq1}
\pt_t \bar{\al} = \bar{\al} G_1(\bar{\al}, \bar{m}^2) \, ,
\\
&&
\label{eq:floweq2}
\pt_t \bar{m}^2 = \bar{\al} H_1(\bar{\al}, \bar{m}^2) 
+ \bar{m}^2 H_2(\bar{\al}, \bar{m}^2) \, ,
\eea
where for simplicity we do not write the full expression for 
$G_1$, $H_1$ and $H_2$. These are complicated and lengthy functions 
of dimensionless coupling parameters ($\bar{\al}$ and $\bar{m}^2$),
and depends on endomorphism parameters ($\g_1$, $\g_2$
and $\g_3$). This simple form of the dimensionless beta-functions 
explicitly shows the occurrence of Gaussian fixed point 
($\bar{\al}=0$ and $\bar{m}^2 = 0$). Moreover, one can use 
these to write the beta-function of dimensionful 
couplings $\al$ and $m^2$, which gets appropriately modified 
by extra terms. These are given by,
\beq
\label{eq:dimfulBT}
\pt\al = \al (2 + G_1(\bar{\al}, \bar{m}^2)) \, ,
\hspace{5mm}
\pt_t m^2 = 2 m^2 + \al H_1(\bar{\al}, \bar{m}^2) 
+ m^2 H_2(\bar{\al}, \bar{m}^2) \, .
\eeq
These dimensionful beta function shows that 
$\al=0$ and $m^2=0$ is a fixed point of the theory. 
The occurrence of this gaussian fixed point is a robust feature.
It is independent of truncation, endomorphism parameters 
and scheme. However the form of 
$G_1$, $H_1$ and $H_2$ is truncation dependent. 
By this we mean that had we extracted the beta-functions
not from coefficient of $\bar{\phi}^2$ and $\bar{\phi}^3$, 
but from higher powers then we still get the Gaussian 
fixed point. 

To compute the stability matrix we use the expression 
given in eq. (\ref{eq:linmat}). For the beta-function given in 
eq. (\ref{eq:floweq1} \& \ref{eq:floweq2}), 
the entries of the stability matrix are given by
(if we take $g_1 = \bar{\al}$ and $g_2 = \bar{m}^2$)
\bea
\label{eq:stabmat}
&&
M_{11} = G_1 + \bar{\al} \frac{\pt G_1}{\pt \bar{\al} }\, , 
\hspace{29mm}
M_{12} = \bar{\al} \frac{\pt G_1}{\pt \bar{m}^2} \, ,
\notag \\
&&
M_{21} = H_1 + \bar{\al} \frac{\pt H_1}{\pt \bar{\al}}
+ \bar{m}^2 \frac{\pt H_2}{\pt \bar{\al}} \, ,
\hspace{10mm}
M_{22} = \bar{\al} \frac{\pt H_1}{\pt \bar{m}^2}
+ H_2 + \bar{m}^2 \frac{\pt H_2}{\pt \bar{m}^2} \, .
\eea
Near the gaussian fixed point ($\bar{\al}=0$ and $\bar{m}^2 = 0$), 
$G_1$, $H_1$ and $H_2$ have the following expansion
for the case of $T$-models,
\bea
\label{eq:GHHexp1}
G_1 &=& -2 + \left(\frac{12}{5} + \frac{339\sqrt{3}}{160} \bar{M}_P \right) \bar{m}^2 
- \frac{10816\sqrt{3} + 137061 \bar{M}_P}{1080 \bar{M}_P^3} \bar{\al} + \cdots \, ,
\\
\label{eq:GHHexp2}
H_1 &=& -\frac{1352}{9 \sqrt{3}\bar{M}_P^3} 
- \frac{481312 + 1223619 \sqrt{3} \bar{M}_P}{1215\bar{M}_P^6} \bar{\al} + \cdots \, ,
\\
H_2 &=& -6 + \frac{12544 \sqrt{3} -372249 \bar{M}_P}{1080 \bar{M}_P^3} \bar{\al}
+ \left( \frac{24}{5} + \frac{933\sqrt{3}\bar{M}_P}{160} \right) \bar{m}^2 
+\cdots \, ,
\eea
where $\bar{M}_P = \mu^{-2} M_P$ is the dimensionless 
Planck mass. One can use this expansion to compute the entries 
of the stability matrix at the Gaussian fixed point. This is given by,
\beq
\label{eq:matfp0}
M = \left(
\begin{array}{c c}
-2 & 0 \\
1352/(9\sqrt{3}\bar{M}_P^3) & -6 
\end{array}
\right) \, .
\eeq
The eigenvalues of this matrix are $e_1=-6$ and $e_2=-2$ with eigenvectors 
$v_1=\{0,1\}^T$ and $v_2 = \{9\sqrt{3}\bar{M}_P^3/338, 1\}^T$ respectively. 
This implies that the gaussian fixed point is a stable point
along both the eigen-directions. Solving the flow of dimensionless 
parameters around this gaussian fixed point it is seen that
\beq
\label{eq:flow0}
\bar{\al}(t) = d_1 e^{-2t} \, ,
\hspace{10mm}
\bar{m}^2(t) = \frac{d_1 338 e^{-6t} (e^{4t} -1)}{9\sqrt{3}\bar{M}_P^3}
+ d_2 e^{-6t} \, ,
\eeq
where $d_1$ and $d_2$ are constants of integration.
This means that in UV (as $t\to\infty$) 
$\bar{\al}$ and $\bar{m}^2$ vanishes. For the corresponding 
dimensionful parameters one can look at dimensionful 
beta-functions given in eq. (\ref{eq:dimfulBT}). These 
clearly show that $\al=0$ is a fixed point. 
It should be noted that as $\al\to0$, 
then the reconstructed $F(R)$ approaches the $R^2$ gravity, 
indicating the stable attractor behaviour of the $R^2$ inflation.

%%%%%%%%%%%%%%%%%%%%%%%%%%%%%%%%%%%%%%%%%%%%%%%
\section{Conclusions}
\label{conc}
%%%%%%%%%%%%%%%%%%%%%%%%%%%%%%%%%%%%%%%%%%%%%%%

In this short paper we have outlined a recipe for the $F(R)$ reconstruction 
process for any given generic potential in the Einstein frame in arbitrary 
space-time dimensions. It is seen that while for simple potentials 
it may be possible to explicitly work out the expression of $F(R)$,
in most cases of $V(\phi)$ one can only express $F(R)$ 
only in parametric form where both $F$ and $R$ are function 
of $\phi$. Analysis of this reconstructed $F(R)$ shows that 
for large $\phi$ (which correspond to large $R$), $F(R) \sim R^2$
in four space-time dimensions. This we explicitly saw in case of 
potentials of the form $\phi^n$, exponential form and $\al$-attractors.
For generic potentials we realised that if $V^\prime/V \to 0$ 
for large $\phi$, then the reconstructed $F(R)$ in 
four space-time dimensions approaches $R^2$ gravity. 
Interestingly in the case of $\al$-attractor this behaviour 
happens for all $\phi$ when $\al\to0$. It therefore becomes 
evident that $R^2$ gravity has a special status in 
four space-time dimensions as an attractor. This classical 
observation has also been made in previous works on 
$F(R)$ reconstruction 
\cite{Rinaldi2014,Bamba2014_1, Bamba2014_2,Pizza2014,Broy2014,Myrzakulov2015}.

In the second part of this paper we set to investigate the quantum theories 
of attractor models with the aim to see if $\al=0$ can be seen as 
a special point in the renormalisation group trajectories. Here 
we make use of non-perturbative renormalisation group flows 
of $F(R)$ models of gravity which have already been computed in 
past in arbitrary dimensions \cite{Ohta20151,Ohta20152,Falls20161,Falls20162}. 
This original equation was written 
for the case where $F$ is an explicit function of $R$ and does not has 
a parametric form. However, it is easy to translate it to the case of 
parametric form by making use of chain rule to express derivatives. 
This allows us to compute the RG flow of reconstructed $F(R)$
in the parametric form. The non-perturbative flow equations for the 
dimensionless parameters $\bar{\al}$ and $\bar{m}^2$ are 
extracted from the non-perturbative RG equation by projecting 
over the required operators. The flow equations clearly show the 
existence of UV stable gaussian fixed point at $\bar{\al}=0$ and $\bar{m}^2=0$.
This is a scheme independent result, indicating the 
the robustness of this fixed point. The RG flow 
of the corresponding dimensionful couplings 
have the same features, where $\al=0$  
while $m^2$ approaches a nonzero value. These are the main 
results of the paper. 

It is worth exploring further the meaning of running of the $\al$-parameter 
during the cosmic evolution. In this RG study $\al$ is a function of 
RG time $t=\ln(\mu/\mu_0)$. In usual flat space-time quantum field theory 
$\mu$ is usually the external momenta of the interacting particles.
In the case of cosmology (and inflation) one can relate 
$\mu$ to the Hubble parameter (or a function of Hubble scale)
\cite{Bonanno2010,Contillo2010,Bonanno2017}. 
Under such a identification of $\mu$, it is worth investigating the effect of the 
running of $\al$-parameter on the cosmic evolution. 
This will be presented in the followup paper. 

The vanishing of $\al$ in UV is a desirable and welcoming feature 
as inflationary models require $\al$ to be small in order to have 
a good inflation. Such a feature is a natural outcome in the context of 
quantum theory where the RG running of parameter $\al$ and its 
UV fixed point keep it small in the inflationary regime. This also supports 
the $R^2$ gravity type models, as via the $F(R)$ reconstruction procedure 
it is seen that the small $\al$ theory tend to higher-derivative gravity models
(similar conclusions were also reached in \cite{Copeland2013} where the 
existence of new nontrivial UV fixed point resolves issue of 
initial conditions). Such higher-derivative gravity models have been known to be good candidates 
for quantum field theory of gravity as they are UV renormalizable 
\cite{Stelle1976} and have been shown to be unitary 
\cite{Narain2011,Narain2012} where ghosts are 
avoided for a large domain of coupling parameter space. 
These models of gravity also offers a favoured realisation 
of inflation. If such models of gravity can be consistently 
extended to explain late time acceleration of universe then 
the full theory becomes a very good model to explain both 
UV and IR physics. Such an attempt has been recently made 
in \cite{Elizalde2017mrn}, where an exponential form of 
$F(R)$ has been added to include dark-energy. This model gives 
us guidelines along which a consistent theory unifying high and low
energy physics should be constructed.

%%%%%%%%%%%%%%%%%%%%%%%%%%%%%%%%%%%%%%%%%%%%%%%
\bigskip
\centerline{\bf Acknowledgements} 
%%%%%%%%%%%%%%%%%%%%%%%%%%%%%%%%%%%%%%%%%%%%%%%

I am thankful to Prof. N. Ohta for enlightening discussions at early stages of the work. 
I am grateful to Prof. Tianjun Li for encouragement and support during the course of this work.  
I am grateful to Nick Houston for discussions and carefully reading the manuscript. 
I would also like to thank Nirmalya Kajuri for discussions and support 
during the course of work.

\appendix

%%%%%%%%%%%%%%%%%%%%%%%%%%%%%%%%%%%%%%%%%%%%%%%
\section{$c_i$'s}
\label{ci}
%%%%%%%%%%%%%%%%%%%%%%%%%%%%%%%%%%%%%%%%%%%%%%%

Here we write the functional form of the $c_i$'s that appear in the 
non-perturbative flow equation (\ref{eq:frgFR}) of the dimensionless $\bar{F}(r)$ 
\cite{Ohta20151,Ohta20152}.
\bea
\label{eq:cifunc}
c_1 &=& \frac{5}{3456\pi^2} (6+(6\g_1-1)r)(6+(6\g_1+1)r)(3+(3\g_1-2)r) ,
\\
c_2 &=& \frac{5}{3456\pi^2} (6+(6\g_1-1)r)[
144 + 9(20\g_1-3)r 
+ 2(6\g_1+1)(3\g_1-2)r^2] ,
\\
c_3 &=& \frac{1}{2304\pi^2} (2+(2\g_2+3)r)(3+(3\g_2-1)r)(6+(6\g_2-5)r) ,
\\
c_4 &=& \frac{1}{256\pi^2}  (2+(2\g_2-1)r)(12+(12\g_2+11)r) ,
\\
c_5 &=& \frac{1}{32\pi^2} \left[12+ 3(8\g_3+1) r +(12\g_3^2 + 3\g_3 -\frac{19}{6})r^2\right] .
\eea
%

%%%%%%%%%%%%%%%%%%%%%%%%%%%%%%%%%%%%%%%%%%%%%%%%%%%%%%%%%%%%%%%%%%%%

%%%%%%%%%%%%%%%%%%%%%%%%%%%%%%%%%%%%%%%%%%%%%%%%%%%%%%%%%%%%%%%%%%%%


\begin{thebibliography}{99} 
%
%
%\cite{Guth:1980zm}
\bibitem{Guth1980} 
  A.~H.~Guth,
  ``The Inflationary Universe: A Possible Solution to the Horizon and Flatness Problems,''
  Phys.\ Rev.\ D {\bf 23}, 347 (1981).
  %%CITATION = PHRVA,D23,347;%%
  %4938 citations counted in INSPIRE as of 23 Aug 2014
%
%
%\cite{Starobinsky:1980te}
\bibitem{Starobinsky1980} 
  A.~A.~Starobinsky,
  ``A New Type of Isotropic Cosmological Models Without Singularity,''
  Phys.\ Lett.\  {\bf 91B}, 99 (1980).
  doi:10.1016/0370-2693(80)90670-X
  %%CITATION = doi:10.1016/0370-2693(80)90670-X;%%
  %3245 citations counted in INSPIRE as of 22 Jul 2017
%
%
%\cite{Linde:1981mu}
\bibitem{Linde1981} 
  A.~D.~Linde,
  ``A New Inflationary Universe Scenario: A Possible Solution of the Horizon, Flatness, 
  Homogeneity, Isotropy and Primordial Monopole Problems,''
  Phys.\ Lett.\ B {\bf 108}, 389 (1982).
  %%CITATION = PHLTA,B108,389;%%
  %2971 citations counted in INSPIRE as of 23 Aug 2014
%
%
%\cite{Linde:1983gd}
\bibitem{Linde1983} 
  A.~D.~Linde,
  ``Chaotic Inflation,''
  Phys.\ Lett.\  {\bf 129B}, 177 (1983).
  doi:10.1016/0370-2693(83)90837-7
  %%CITATION = doi:10.1016/0370-2693(83)90837-7;%%
  %2406 citations counted in INSPIRE as of 22 Jul 2017
%
%
%\cite{Starobinsky:1982ee}
\bibitem{Starobinsky1982} 
  A.~A.~Starobinsky,
  ``Dynamics of Phase Transition in the New Inflationary Universe Scenario and Generation of Perturbations,''
  Phys.\ Lett.\ B {\bf 117}, 175 (1982).
  %%CITATION = PHLTA,B117,175;%%
  %1596 citations counted in INSPIRE as of 23 Aug 2014
%
%
%\cite{Martin:2013tda}
\bibitem{Martin2013} 
  J.~Martin, C.~Ringeval and V.~Vennin,
  ``Encyclop¾dia Inflationaris,''
  Phys.\ Dark Univ.\  {\bf 5-6}, 75 (2014)
  doi:10.1016/j.dark.2014.01.003
  [arXiv:1303.3787 [astro-ph.CO]].
  %%CITATION = doi:10.1016/j.dark.2014.01.003;%%
  %354 citations counted in INSPIRE as of 22 Jul 2017
%
%
%\cite{Sotiriou:2008rp}
\bibitem{Sotiriou2008}
  T.~P.~Sotiriou and V.~Faraoni,
  ``f(R) Theories Of Gravity,''
  Rev.\ Mod.\ Phys.\  {\bf 82} (2010) 451
  doi:10.1103/RevModPhys.82.451
  [arXiv:0805.1726 [gr-qc]].
  %%CITATION = doi:10.1103/RevModPhys.82.451;%%
  %1801 citations counted in INSPIRE as of 23 Jul 2017
%
%
%\cite{DeFelice:2010aj}
\bibitem{DeFelice2010} 
  A.~De Felice and S.~Tsujikawa,
  ``f(R) theories,''
  Living Rev.\ Rel.\  {\bf 13}, 3 (2010)
  doi:10.12942/lrr-2010-3
  [arXiv:1002.4928 [gr-qc]].
  %%CITATION = doi:10.12942/lrr-2010-3;%%
  %1314 citations counted in INSPIRE as of 23 Jul 2017
%
%
%\cite{Maeda:1987xf}
\bibitem{Maeda1987}
  K.~i.~Maeda,
  ``Inflation as a Transient Attractor in R**2 Cosmology,''
  Phys.\ Rev.\ D {\bf 37} (1988) 858.
  doi:10.1103/PhysRevD.37.858
  %%CITATION = doi:10.1103/PhysRevD.37.858;%%
  %126 citations counted in INSPIRE as of 27 Apr 2017
%
%
%\cite{Barrow:1988xh}
\bibitem{Barrow1988} 
  J.~D.~Barrow and S.~Cotsakis,
  ``Inflation and the Conformal Structure of Higher Order Gravity Theories,''
  Phys.\ Lett.\ B {\bf 214}, 515 (1988).
  doi:10.1016/0370-2693(88)90110-4
  %%CITATION = doi:10.1016/0370-2693(88)90110-4;%%
  %353 citations counted in INSPIRE as of 27 Apr 2017
%
%
%\cite{Rinaldi:2014gua}
\bibitem{Rinaldi2014}
  M.~Rinaldi, G.~Cognola, L.~Vanzo and S.~Zerbini,
  ``Reconstructing the inflationary $f(R)$ from observations,''
  JCAP {\bf 1408} (2014) 015
  doi:10.1088/1475-7516/2014/08/015
  [arXiv:1406.1096 [gr-qc]].
  %%CITATION = doi:10.1088/1475-7516/2014/08/015;%%
  %37 citations counted in INSPIRE as of 22 Jul 2017
%
%
%\cite{Bamba:2014daa}
\bibitem{Bamba2014_1} 
  K.~Bamba, S.~Nojiri and S.~D.~Odintsov,
  ``Reconstruction of scalar field theories realizing inflation consistent with the Planck and BICEP2 results,''
  Phys.\ Lett.\ B {\bf 737}, 374 (2014)
  doi:10.1016/j.physletb.2014.09.014
  [arXiv:1406.2417 [hep-th]].
  %%CITATION = doi:10.1016/j.physletb.2014.09.014;%%
  %35 citations counted in INSPIRE as of 22 Jul 2017
%
%
%\cite{Bamba:2014wda}
\bibitem{Bamba2014_2} 
  K.~Bamba, S.~Nojiri, S.~D.~Odintsov and D.~S‡ez-G—mez,
  ``Inflationary universe from perfect fluid and $F(R)$ gravity and its comparison with observational data,''
  Phys.\ Rev.\ D {\bf 90}, 124061 (2014)
  doi:10.1103/PhysRevD.90.124061
  [arXiv:1410.3993 [hep-th]].
  %%CITATION = doi:10.1103/PhysRevD.90.124061;%%
  %61 citations counted in INSPIRE as of 22 Jul 2017
%
%
%\cite{Pizza:2014rqa}
\bibitem{Pizza2014} 
  L.~Pizza,
  ``Numerical approach to model independently reconstruct $f(R)$ functions through cosmographic data,''
  Phys.\ Rev.\ D {\bf 91}, no. 12, 124048 (2015)
  doi:10.1103/PhysRevD.91.124048
  [arXiv:1411.5348 [astro-ph.CO]].
  %%CITATION = doi:10.1103/PhysRevD.91.124048;%%
  %2 citations counted in INSPIRE as of 22 Jul 2017
%
%
%\cite{Broy:2014xwa}
\bibitem{Broy2014}
  B.~J.~Broy, F.~G.~Pedro and A.~Westphal,
  ``Disentangling the $f(R)$ - Duality,''
  JCAP {\bf 1503} (2015) no.03,  029
  doi:10.1088/1475-7516/2015/03/029
  [arXiv:1411.6010 [hep-th]].
  %%CITATION = doi:10.1088/1475-7516/2015/03/029;%%
  %18 citations counted in INSPIRE as of 22 Jul 2017
%
%
%\cite{Myrzakulov:2015fra}
\bibitem{Myrzakulov2015} 
  R.~Myrzakulov, L.~Sebastiani and S.~Zerbini,
  ``Reconstruction of Inflation Models,''
  Eur.\ Phys.\ J.\ C {\bf 75}, no. 5, 215 (2015)
  doi:10.1140/epjc/s10052-015-3443-4
  [arXiv:1502.04432 [gr-qc]].
  %%CITATION = doi:10.1140/epjc/s10052-015-3443-4;%%
  %16 citations counted in INSPIRE as of 22 Jul 2017
%
%
%\cite{Kallosh:2015lwa}
\bibitem{Kallosh2015} 
  R.~Kallosh and A.~Linde,
  ``Planck, LHC, and $\alpha$-attractors,''
  Phys.\ Rev.\ D {\bf 91}, 083528 (2015)
  doi:10.1103/PhysRevD.91.083528
  [arXiv:1502.07733 [astro-ph.CO]].
  %%CITATION = doi:10.1103/PhysRevD.91.083528;%%
  %61 citations counted in INSPIRE as of 22 Jul 2017
%
%
%\cite{Linde:2015uga}
\bibitem{Linde2015} 
  A.~Linde,
  ``Single-field $\alpha$-attractors,''
  JCAP {\bf 1505}, 003 (2015)
  doi:10.1088/1475-7516/2015/05/003
  [arXiv:1504.00663 [hep-th]].
  %%CITATION = doi:10.1088/1475-7516/2015/05/003;%%
  %43 citations counted in INSPIRE as of 22 Jul 2017
%
%
%\cite{Carrasco:2015rva}
\bibitem{Carrasco2015_1} 
  J.~J.~M.~Carrasco, R.~Kallosh and A.~Linde,
  ``Cosmological Attractors and Initial Conditions for Inflation,''
  Phys.\ Rev.\ D {\bf 92}, no. 6, 063519 (2015)
  doi:10.1103/PhysRevD.92.063519
  [arXiv:1506.00936 [hep-th]].
  %%CITATION = doi:10.1103/PhysRevD.92.063519;%%
  %45 citations counted in INSPIRE as of 22 Jul 2017
%
%
%\cite{Carrasco:2015pla}
\bibitem{Carrasco2015_2} 
  J.~J.~M.~Carrasco, R.~Kallosh and A.~Linde,
  ``$\alpha $-Attractors: Planck, LHC and Dark Energy,''
  JHEP {\bf 1510}, 147 (2015)
  doi:10.1007/JHEP10(2015)147
  [arXiv:1506.01708 [hep-th]].
  %%CITATION = doi:10.1007/JHEP10(2015)147;%%
  %56 citations counted in INSPIRE as of 22 Jul 2017
%
%
%\cite{Kallosh:2016gqp}
\bibitem{Kallosh2016} 
  R.~Kallosh and A.~Linde,
  ``Cosmological Attractors and Asymptotic Freedom of the Inflaton Field,''
  JCAP {\bf 1606}, no. 06, 047 (2016)
  doi:10.1088/1475-7516/2016/06/047
  [arXiv:1604.00444 [hep-th]].
  %%CITATION = doi:10.1088/1475-7516/2016/06/047;%%
  %20 citations counted in INSPIRE as of 22 Jul 2017
%
%
%\cite{Linde:2016uec}
\bibitem{Linde2016} 
  A.~Linde,
  ``Random Potentials and Cosmological Attractors,''
  JCAP {\bf 1702}, no. 02, 028 (2017)
  doi:10.1088/1475-7516/2017/02/028
  [arXiv:1612.04505 [hep-th]].
  %%CITATION = doi:10.1088/1475-7516/2017/02/028;%%
  %5 citations counted in INSPIRE as of 22 Jul 2017
%
%
%\cite{Pinhero:2017lni}
\bibitem{Pinhero2017} 
  T.~Pinhero and S.~Pal,
  ``Non-canonical Conformal Attractors for Single Field Inflation,''
  arXiv:1703.07165 [hep-th].
  %%CITATION = ARXIV:1703.07165;%%
  %1 citations counted in INSPIRE as of 22 Jul 2017
%
%
%\cite{Kallosh:2013wya}
\bibitem{Kallosh2013} 
  R.~Kallosh and A.~Linde,
  ``Superconformal generalization of the chaotic inflation model $\frac{\lambda}{4} \phi^{4} - \frac{\xi}{2} \phi^{2}R$,''
  JCAP {\bf 1306}, 027 (2013)
  doi:10.1088/1475-7516/2013/06/027
  [arXiv:1306.3211 [hep-th]].
  %%CITATION = doi:10.1088/1475-7516/2013/06/027;%%
  %74 citations counted in INSPIRE as of 25 Sep 2017
%
%
%\cite{Kallosh:2013maa}
\bibitem{Kallosh2013maa} 
  R.~Kallosh and A.~Linde,
  ``Non-minimal Inflationary Attractors,''
  JCAP {\bf 1310}, 033 (2013)
  doi:10.1088/1475-7516/2013/10/033
  [arXiv:1307.7938 [hep-th]].
  %%CITATION = doi:10.1088/1475-7516/2013/10/033;%%
  %72 citations counted in INSPIRE as of 25 Sep 2017
%
%
%\cite{Elizalde:2015nya}
\bibitem{Elizalde2015} 
  E.~Elizalde, S.~D.~Odintsov, E.~O.~Pozdeeva and S.~Y.~Vernov,
  ``Cosmological attractor inflation from the RG-improved Higgs sector of finite gauge theory,''
  JCAP {\bf 1602}, no. 02, 025 (2016)
  doi:10.1088/1475-7516/2016/02/025
  [arXiv:1509.08817 [gr-qc]].
  %%CITATION = doi:10.1088/1475-7516/2016/02/025;%%
  %15 citations counted in INSPIRE as of 25 Sep 2017
%
%
%\cite{Choudhury:2017cos}
\bibitem{Choudhury2017} 
  S.~Choudhury,
  ``COSMOS-$e'$- soft Higgsotic attractors,''
  Eur.\ Phys.\ J.\ C {\bf 77}, no. 7, 469 (2017)
  doi:10.1140/epjc/s10052-017-5001-8
  [arXiv:1703.01750 [hep-th]].
  %%CITATION = doi:10.1140/epjc/s10052-017-5001-8;%%
  %2 citations counted in INSPIRE as of 25 Sep 2017
%
%
%\cite{Mishra:2017ehw}
\bibitem{Mishra2017} 
  S.~S.~Mishra, V.~Sahni and Y.~Shtanov,
  ``Sourcing Dark Matter and Dark Energy from $\alpha$-attractors,''
  JCAP {\bf 1706}, no. 06, 045 (2017)
  doi:10.1088/1475-7516/2017/06/045
  [arXiv:1703.03295 [gr-qc]].
  %%CITATION = doi:10.1088/1475-7516/2017/06/045;%%
  %3 citations counted in INSPIRE as of 01 Aug 2017
%
%
%\cite{Odintsov:2016vzz}
\bibitem{Odintsov2016} 
  S.~D.~Odintsov and V.~K.~Oikonomou,
  ``Inflationary $\alpha$-attractors from $F(R)$ gravity,''
  Phys.\ Rev.\ D {\bf 94}, no. 12, 124026 (2016)
  doi:10.1103/PhysRevD.94.124026
  [arXiv:1612.01126 [gr-qc]].
  %%CITATION = doi:10.1103/PhysRevD.94.124026;%%
  %9 citations counted in INSPIRE as of 22 Jul 2017
%
%
%\cite{Bhattacharya:2017uwi}
\bibitem{Bhattacharya2017} 
  S.~Bhattacharya, K.~Das and K.~Dutta,
  ``Attractor Models in Scalar-Tensor Theories of Inflation,''
  arXiv:1706.07934 [gr-qc].
  %%CITATION = ARXIV:1706.07934;%%
  %1 citations counted in INSPIRE as of 22 Jul 2017
%
%
%\cite{Fumagalli:2016sof}
\bibitem{Fumagalli2016} 
  J.~Fumagalli,
  ``Renormalization Group independence of Cosmological Attractors,''
  Phys.\ Lett.\ B {\bf 769}, 451 (2017)
  doi:10.1016/j.physletb.2017.04.017
  [arXiv:1611.04997 [hep-th]].
  %%CITATION = doi:10.1016/j.physletb.2017.04.017;%%
  %1 citations counted in INSPIRE as of 22 Aug 2017
%
%
%\cite{Miranda:2017juz}
\bibitem{Miranda2017} 
  T.~Miranda, J.~C.~Fabris and O.~F.~Piattella,
  ``Reconstructing a $f(R)$ theory from the $\alpha$-Attractors,''
  arXiv:1707.06457 [gr-qc].
  %%CITATION = ARXIV:1707.06457;%%
%
%
%\cite{Wetterich:1992yh}
\bibitem{Wetterich1992}
  C.~Wetterich,
  ``Exact evolution equation for the effective potential,''
  Phys.\ Lett.\ B {\bf 301} (1993) 90.
  doi:10.1016/0370-2693(93)90726-X
  %%CITATION = doi:10.1016/0370-2693(93)90726-X;%%
  %1091 citations counted in INSPIRE as of 23 Jul 2017
%
%
%\cite{Machado:2007ea}
\bibitem{Machado2007} 
  P.~F.~Machado and F.~Saueressig,
  ``On the renormalization group flow of f(R)-gravity,''
  Phys.\ Rev.\ D {\bf 77}, 124045 (2008)
  doi:10.1103/PhysRevD.77.124045
  [arXiv:0712.0445 [hep-th]].
  %%CITATION = doi:10.1103/PhysRevD.77.124045;%%
  %166 citations counted in INSPIRE as of 23 Jul 2017
%
%
%\cite{Codello:2007bd}
\bibitem{Codello2007} 
  A.~Codello, R.~Percacci and C.~Rahmede,
  ``Ultraviolet properties of f(R)-gravity,''
  Int.\ J.\ Mod.\ Phys.\ A {\bf 23}, 143 (2008)
  doi:10.1142/S0217751X08038135
  [arXiv:0705.1769 [hep-th]].
  %%CITATION = doi:10.1142/S0217751X08038135;%%
  %183 citations counted in INSPIRE as of 23 Jul 2017
%
%
%\cite{Codello:2008vh}
\bibitem{Codello2008} 
  A.~Codello, R.~Percacci and C.~Rahmede,
  ``Investigating the Ultraviolet Properties of Gravity with a Wilsonian Renormalization Group Equation,''
  Annals Phys.\  {\bf 324}, 414 (2009)
  doi:10.1016/j.aop.2008.08.008
  [arXiv:0805.2909 [hep-th]].
  %%CITATION = doi:10.1016/j.aop.2008.08.008;%%
  %307 citations counted in INSPIRE as of 23 Jul 2017
%
%
%\cite{Ohta:2015efa}
\bibitem{Ohta20151} 
  N.~Ohta, R.~Percacci and G.~P.~Vacca,
  ``Flow equation for $f(R)$ gravity and some of its exact solutions,''
  Phys.\ Rev.\ D {\bf 92}, no. 6, 061501 (2015)
  doi:10.1103/PhysRevD.92.061501
  [arXiv:1507.00968 [hep-th]].
  %%CITATION = doi:10.1103/PhysRevD.92.061501;%%
  %34 citations counted in INSPIRE as of 23 Jul 2017
%
%
%\cite{Ohta:2015fcu}
\bibitem{Ohta20152} 
  N.~Ohta, R.~Percacci and G.~P.~Vacca,
  ``Renormalization Group Equation and scaling solutions for f(R) gravity in exponential parametrization,''
  Eur.\ Phys.\ J.\ C {\bf 76}, no. 2, 46 (2016)
  doi:10.1140/epjc/s10052-016-3895-1
  [arXiv:1511.09393 [hep-th]].
  %%CITATION = doi:10.1140/epjc/s10052-016-3895-1;%%
  %21 citations counted in INSPIRE as of 23 Jul 2017
%
%
%\cite{Falls:2016wsa}
\bibitem{Falls20161} 
  K.~Falls, D.~F.~Litim, K.~Nikolakopoulos and C.~Rahmede,
  ``On de Sitter solutions in asymptotically safe $f(R)$ theories,''
  arXiv:1607.04962 [gr-qc].
  %%CITATION = ARXIV:1607.04962;%%
  %10 citations counted in INSPIRE as of 23 Jul 2017
%
%
%\cite{Falls:2016msz}
\bibitem{Falls20162} 
  K.~Falls and N.~Ohta,
  ``Renormalization Group Equation for $f(R)$ gravity on hyperbolic spaces,''
  Phys.\ Rev.\ D {\bf 94}, no. 8, 084005 (2016)
  doi:10.1103/PhysRevD.94.084005
  [arXiv:1607.08460 [hep-th]].
  %%CITATION = doi:10.1103/PhysRevD.94.084005;%%
  %10 citations counted in INSPIRE as of 12 Jun 2017
%
%
%\cite{Niedermaier:2006wt}
\bibitem{Niedermaier2006} 
  M.~Niedermaier and M.~Reuter,
  ``The Asymptotic Safety Scenario in Quantum Gravity,''
  Living Rev.\ Rel.\  {\bf 9}, 5 (2006).
  doi:10.12942/lrr-2006-5
  %%CITATION = doi:10.12942/lrr-2006-5;%%
  %345 citations counted in INSPIRE as of 23 Jul 2017
%
%
%\cite{Percacci:2007sz}
\bibitem{Percacci2007} 
  R.~Percacci,
  ``Asymptotic Safety,''
  In *Oriti, D. (ed.): Approaches to quantum gravity* 111-128
  [arXiv:0709.3851 [hep-th]].
  %%CITATION = ARXIV:0709.3851;%%
  %198 citations counted in INSPIRE as of 29 Jan 2017
%
%
%\cite{Kamenshchik:2014waa}
\bibitem{Kamenshchik2014} 
  A.~Y.~Kamenshchik and C.~F.~Steinwachs,
  ``Question of quantum equivalence between Jordan frame and Einstein frame,''
  Phys.\ Rev.\ D {\bf 91}, no. 8, 084033 (2015)
  doi:10.1103/PhysRevD.91.084033
  [arXiv:1408.5769 [gr-qc]].
  %%CITATION = doi:10.1103/PhysRevD.91.084033;%%
  %34 citations counted in INSPIRE as of 24 Jul 2017
%
%
%\cite{Banerjee:2016lco}
\bibitem{Banerjee2016}
  N.~Banerjee and B.~Majumder,
  ``A question mark on the equivalence of Einstein and Jordan frames,''
  Phys.\ Lett.\ B {\bf 754} (2016) 129
  doi:10.1016/j.physletb.2016.01.022
  [arXiv:1601.06152 [gr-qc]].
  %%CITATION = doi:10.1016/j.physletb.2016.01.022;%%
  %9 citations counted in INSPIRE as of 24 Jul 2017
%
%
%\cite{Pandey:2016unk}
\bibitem{Pandey2016} 
  S.~Pandey and N.~Banerjee,
  ``Equivalence of Jordan and Einstein frames at the quantum level,''
  Eur.\ Phys.\ J.\ Plus {\bf 132}, no. 3, 107 (2017)
  doi:10.1140/epjp/i2017-11385-0
  [arXiv:1610.00584 [gr-qc]].
  %%CITATION = doi:10.1140/epjp/i2017-11385-0;%%
  %4 citations counted in INSPIRE as of 24 Jul 2017
%
%
%\cite{Bonanno:2010bt}
\bibitem{Bonanno2010} 
  A.~Bonanno, A.~Contillo and R.~Percacci,
  ``Inflationary solutions in asymptotically safe f(R) theories,''
  Class.\ Quant.\ Grav.\  {\bf 28}, 145026 (2011)
  doi:10.1088/0264-9381/28/14/145026
  [arXiv:1006.0192 [gr-qc]].
  %%CITATION = doi:10.1088/0264-9381/28/14/145026;%%
  %54 citations counted in INSPIRE as of 30 Jul 2017
%
%
%\cite{Contillo:2010ju}
\bibitem{Contillo2010} 
  A.~Contillo,
  ``Evolution of cosmological perturbations in an RG-driven inflationary scenario,''
  Phys.\ Rev.\ D {\bf 83}, 085016 (2011)
  doi:10.1103/PhysRevD.83.085016
  [arXiv:1011.4618 [gr-qc]].
  %%CITATION = doi:10.1103/PhysRevD.83.085016;%%
  %12 citations counted in INSPIRE as of 30 Jul 2017
%
%
%\cite{Bonanno:2017pkg}
\bibitem{Bonanno2017} 
  A.~Bonanno and F.~Saueressig,
  ``Asymptotically safe cosmology Ð A status report,''
  Comptes Rendus Physique {\bf 18}, 254
  doi:10.1016/j.crhy.2017.02.002
  [arXiv:1702.04137 [hep-th]].
  %%CITATION = doi:10.1016/j.crhy.2017.02.002;%%
  %10 citations counted in INSPIRE as of 30 Jul 2017
%
%
%\cite{Copeland:2013vva}
\bibitem{Copeland2013} 
  E.~J.~Copeland, C.~Rahmede and I.~D.~Saltas,
  ``Asymptotically Safe Starobinsky Inflation,''
  Phys.\ Rev.\ D {\bf 91}, no. 10, 103530 (2015)
  doi:10.1103/PhysRevD.91.103530
  [arXiv:1311.0881 [gr-qc]].
  %%CITATION = doi:10.1103/PhysRevD.91.103530;%%
  %36 citations counted in INSPIRE as of 25 Sep 2017
%
%
%\cite{Stelle:1976gc}
\bibitem{Stelle1976} 
  K.~S.~Stelle,
  ``Renormalization of Higher Derivative Quantum Gravity,''
  Phys.\ Rev.\ D {\bf 16}, 953 (1977).
  doi:10.1103/PhysRevD.16.953
  %%CITATION = doi:10.1103/PhysRevD.16.953;%%
  %1421 citations counted in INSPIRE as of 25 Sep 2017
%
%
%\cite{Narain:2011gs}
\bibitem{Narain2011} 
  G.~Narain and R.~Anishetty,
  ``Short Distance Freedom of Quantum Gravity,''
  Phys.\ Lett.\ B {\bf 711}, 128 (2012)
  doi:10.1016/j.physletb.2012.03.070
  [arXiv:1109.3981 [hep-th]].
  %%CITATION = doi:10.1016/j.physletb.2012.03.070;%%
  %14 citations counted in INSPIRE as of 25 Sep 2017
%
%
%\cite{Narain:2012nf}
\bibitem{Narain2012} 
  G.~Narain and R.~Anishetty,
  ``Unitary and Renormalizable Theory of Higher Derivative Gravity,''
  J.\ Phys.\ Conf.\ Ser.\  {\bf 405}, 012024 (2012)
  doi:10.1088/1742-6596/405/1/012024
  [arXiv:1210.0513 [hep-th]].
  %%CITATION = doi:10.1088/1742-6596/405/1/012024;%%
  %9 citations counted in INSPIRE as of 25 Sep 2017
%
%
%\cite{Elizalde:2017mrn}
\bibitem{Elizalde2017mrn} 
  E.~Elizalde, S.~D.~Odintsov, L.~Sebastiani and R.~Myrzakulov,
  ``Beyond-one-loop quantum gravity action yielding both inflation and late-time acceleration,''
  Nucl.\ Phys.\ B {\bf 921}, 411 (2017)
  doi:10.1016/j.nuclphysb.2017.06.003
  [arXiv:1706.01879 [gr-qc]].
  %%CITATION = doi:10.1016/j.nuclphysb.2017.06.003;%%
  %2 citations counted in INSPIRE as of 25 Sep 2017
%
%
\end{thebibliography}
\end{document}